\documentclass[10pt,journal,compsoc]{IEEEtran}
\makeatletter
\newcommand{\thickhline}{%
    \noalign {\ifnum 0=`}\fi \hrule height 1pt
    \futurelet \reserved@a \@xhline
}

\ifCLASSOPTIONcompsoc
  \usepackage[nocompress]{cite}
\else
  \usepackage{cite}
\fi
\usepackage[cmex10]{amsmath}
\usepackage{amssymb}
\usepackage{lscape}
\graphicspath{{./eps/}}
\usepackage{url}
\usepackage{amsthm}
\theoremstyle{definition}
\newtheorem{definition}{Definition}
\newtheorem*{remark}{Remark}
\usepackage{algorithm}
\usepackage{algorithmic}
\usepackage{enumerate}
\usepackage{graphicx}
\usepackage{subfig}
\usepackage{placeins} 
\usepackage[acronym]{glossaries}

\makeglossaries

\newacronym{gcd}{GCD}{Greatest Common Divisor}

\newacronym{lcm}{LCM}{Least Common Multiple}

\ifCLASSINFOpdf
\else
\fi

\begin{document}
%
\title{Scaling up for high dimensional and high speed data streams: HSDStream}
%
%
%
%
\author{Irshad~Ahmed,~
        Irfan~Ahmed,~\IEEEmembership{Member,~IEEE,}
        and~Waseem~Shahzad,
\thanks{Irfan Ahmed is with the Department of Computer Engineering, Taif University, Taif,
21974 Saudi Arabia e-mail: i.ahmed@tu.edu.sa.}
\thanks{Irshad Ahmed and Waseem Shahzad are with National University of Computer and Emerging Sciences, Islamabad, Pakistan.}
}


\IEEEtitleabstractindextext{%
\begin{abstract}
This paper presents a novel high speed clustering scheme for high dimensional data streams. Data stream clustering has gained importance in different applications, for example, in network monitoring, intrusion detection, and real-time sensing are few of those. High dimensional stream data is inherently more complex when used for clustering because the evolving nature of the stream data and high dimensionality make it non-trivial. In order to tackle this problem, projected subspace within the high dimensions and limited window sized data per unit of time are used for clustering purpose. We propose a High Speed and Dimensions data stream clustering scheme (HSDStream) which employs exponential moving averages to reduce the size of the memory and speed up the processing of projected subspace data stream. The proposed algorithm has been tested against HDDStream for cluster purity, memory usage, and the cluster sensitivity. Experimental results have been obtained for corrected KDD intrusion detection dataset. These results show that HSDStream outperforms the HDDStream in all performance metrics, especially the memory usage and the processing speed.
\end{abstract}

}

\maketitle

\IEEEdisplaynontitleabstractindextext

%
\IEEEpeerreviewmaketitle

\section{Introduction}\label{sec:introduction}
\IEEEPARstart{T}{he} exponential growth in data mining and clustering is an apparent result of network applications that are becoming part of our daily life. In today's applications, whether they are related to academic, research, finance, business, or military, the evolving data streams are ubiquitous.
Data sources are monotonically increasing from past few decades. Additionally, the technological developments in data sensing systems (sensor networks) have resulted in a real-time data with large number of attributes. The large volume of the data together with its high dimensionality has motivated the research in the area of high dimensional data mining and exploration. Data stream is a form of data that continuously evolves reflecting the real-time variation in volume, dimensionality, and correlation.
In recent years, a large amount of streaming data, such as network flows, wireless sensor networks data and the multimedia streams have been generated. Analyzing and mining of real-time streaming data have become a hot research topic \cite{forestiero_single_2013}, \cite{sim_survey_2013}, \cite{aggarwal_segment-based_2012}. Discovery of the patterns hidden in the streaming data imposes great challenges for cluster formation, especially in high dimensional data.
By definition, a cluster is a collection of objects which are “similar” between them and are “dissimilar” to the objects belonging to other clusters.
 Data stream clustering algorithms are used to get important information from these streams in real-time.
These algorithms search for the clusters that contain streaming objects with a certain degree of similarity across all dimensions. Stream clustering algorithms have special challenges that do not face most other clustering techniques. Storage and time limits are critical for clustering algorithms to perform a fast single-pass over that stream data.
In addition to this, the evolving nature of the stream, requires the clustering algorithm to be highly adaptive to the new patterns. Generally, there are two types of stream clustering algorithms: full dimensional and projected or preferred dimension streaming algorithms. Clustering applications in various domains often have very high-dimensional data; the dimension of the data being in the tens, hundreds or thousands, for example, in network streaming, web mining and bioinformatics, respectively. It is often require to focus on a certain subset of dimensions rather than the full dimension space because it requires less memory and render fast processing. In addition to the high dimensionality, real-time high-speed evolution makes it more intractable. Clustering such high-dimensional high-speed datasets is a contemporary challenge. Clustering algorithms must avoid the curse of dimensionality but at the same time should be computationally efficient.
Some applications that generate data streams include: telecommunication (call records), network operation centers (log information from network entities), financial market (stock exchange), and day to day business (credit card, ATM transactions, etc).
In a high dimensional dataset, among many features  some attributes can be expected to be irrelevant for any given object of interest. Irrelevant attributes can obscure clusters that are clearly visible when we consider only the relevant ‘subspace’ of the dataset. Therefore, clusters may be meaningfully defined by some of the available attributes only.
The irrelevant attributes interfere with the efforts to find targeted clusters. This problem is become more intensive in
streaming data, because it requires a single scan of the data to find the useful attributes for describing a potential cluster for the current object. Moreover, streams are impulsive and the discovered clusters might also evolve over time. High dimensional streaming data clustering is more challenging than the high density or high dimensional data. Among various challenges in clustering high dimensional streaming data \cite{amini_density-based_2014}, following two are the focuses of this paper:
\begin{itemize}
  \item Processing speed: Data streams arrive continuously, which requires fast and real-time response. The clustering algorithm needs to have processing speed (which comes from low complexity) such that it can handle the speed of data streams in the limited time.
  \item Memory usage: Large data streams are generated rapidly which need an unlimited memory. Therefore, the clustering algorithm must be optimized for realistic memory constraints.
\end{itemize}

\emph{Notations}:\\
Vectors and matrices are represented by bold letters, other notations are explained below:\\
\begin{tabular}{c l}
  $\mathbb{R}$ & Set of real numbers \\
  $\mathbb{N}$ & Set of natural numbers  \\
  $\mathcal{C}$ & Dataset\\
  N & Window size \\
  $\epsilon$ & Radius threshold \\
  $\mathcal{D}$ & Dataset used in initialization phases\\
  $\alpha$ & Exponential weighted average constant\\
  $\beta$ & Outlier threshold\\
  $\mu$ &  Number of points threshold\\
  $\xi$ & Variance threshold\\
  $\psi_{j}$  & $j^{th}$ preferred dimension\\
  $\pi$ & Projected dimensionality threshold\\
\end{tabular}

\section{Related Work}
In the last few years many research works have been done on high dimensional data clustering and evolving data streams clustering.
There are extensive research works on clustering algorithms for static datasets \cite{jain_data_2010,kriegel_clustering_2009,amini_density-based_2014} where some of them have been further extended for evolving data streams.  The clusters are formed based on a Euclidean distance function like \emph{k}-means algorithm \cite{macqueen_methods_1967}. \emph{k}-mean clustering splits the $n$ d-dimensional points into $k$ cluster ($k<n$). One of the well-known extensions of \emph{k}-means on data streams is presented by Aggarwal et al. \cite{aggarwal_framework_2003}. They proposed an algorithm called CluStream based on \emph{k}-means for clustering evolving data streams. CluStream introduces an online-offline method for clustering data streams. CluStream clustering idea has been adopted for the majority of data stream clustering algorithms. Aggarwal et al. extended their work in HPStream \cite{aggarwal_framework_2004}, which introduces the projected clustering to data streams. In projected clustering high dimensional stream data has been partitioned based on preferred dimensions instead of full dimensional space. Cao et al. \cite{cao_density-based_2006} use the density-based clustering without projected dimensions in DenStream algorithm.
For streaming data, although a considerable research has tackled the full-space clustering, relatively limited work has been dealt with subspace clustering. These few researches include \cite{aggarwal_framework_2004} HPStream, \cite{ntoutsi_density-based_2012} HDDStream, and \cite{hassani_subspace_2014} SubCMM. A more comprehensive review and classifications are given in survey \cite{nguyen_survey_2014}.
In \cite{ntoutsi_density-based_2012}, authors proposed a density-based projected clustering scheme for high dimensional data streams called HDDStream. HDDStream works in three phases; an initial phase in which initial set of core micro-clusters has been formed, then online core and outlier clusters' maintenance with projected clustering, and finally, an on-demand offline clustering phase. Compared with HPStream which requires the fixed number of clusters, the number of clusters in HDDStream is variably adjusted over time, and the clusters can be of arbitrary shape.
SubCMM suggests a different way for evaluating stream subspace clustering algorithms by making use of available offline subspace clustering algorithms with the streaming environment to handle the errors caused by emerging, moving, or splitting subspace clusters.
A recent, similarity-based Data Stream Classifier (SimC)\cite{mena-torres_similarity-based_2014} introduces an insertion/removal policy that adapts evolving data tendency and maintains a representative, small set of clusters. It uses instance based learning techniques to form adaptive clustering algorithm.
In \cite{forestiero_single_2013} clustering method based on a multi-agent system that uses a decentralized bottom-up self-organizing strategy to group similar data points has been presented. It uses bio-inspired flocking model to eliminate the need of offline clustering. A clustering algorithm for stream data with uncertain attributes has been presented in \cite{jin_efficient_2014}. This scheme works only for low dimensional streaming data. Liu \cite{liu_clustering_2011} developed HSWStream algorithm. It is a data stream clustering algorithm  based on exponential histogram over sliding windows with projected dimensions. Another density-based algorithm D-Stream \cite{wan_density-based_2009} maps each input data into a grid, computes the density of each grid, and forms the clusters using these grids. In \cite{lee_efficiently_2009}, authors proposed a scalable algorithm to trace clusters in a high-dimensional data stream. The proposed scheme transforms the problem of multi-dimensional clustering into that of one-dimensional clustering along with a frequent itemset mining technique. This scheme achieves the scalability on the number of dimensions while sacrificing the accuracy of identified clusters. Bellas et al. \cite{bellas_model-based_2013} presented an online variant of mixture of probabilistic principal component analyzers (MPPCA) to model and cluster the high dimensional high speed data. But to do so, it is necessary to add a classification step at the end of the online MPPCA algorithm to provide the expected clustering. MuDi-Stream \cite{amini_mudi-stream:_2014} is a hybrid grid-based multi-density clustering algorithm with online-offline phases. In the online phase, it keeps summary information of evolving multi-density data stream in the form of core micro-clusters. The offline phase generates the final clusters using an adapted density-based clustering algorithm. The grid-based method is used as an outlier buffer to handle both noises and multi-density data in order to reduce the merging time of clustering. MuDi-Stream is not suitable for high-dimensional data since the number of empty grids increases which requires longer processing time. SE-Stream \cite{huynh_se-stream:_2014} is a standard-deviation based projected clustering method to support high dimensional data streams. It forms clusters within subgroups of dimensions and can detect change in the clustering structure during the progression of data streams. SED-Stream \cite{waiyamai_sed-stream:_2014} is an extension of SE-Stream, in which some selected dimensions are used to represent the clusters to increase the quality of the output clustering. SED-Stream projects any cluster to its discriminative dimensions that are highly relevant to the cluster itself but distinguished from the other clusters. SED-Stream is better than its previous version, SE-Stream, in terms of purity and f-measure. Both SE-Stream and SED-Stream use fading cluster structure ($5-tuple$) of the form similar to in section \ref{sec:ProblemFormulation} definition $0$ with two extra elements. \\
This paper presents HSDStream which introduces a novel tuple structure to summarize the high speed high dimensional data stream. This structure not only speed up the process but also requires less memory. Our clustering technique also modifies weights in some definitions of HDDStream, namely, the micro-cluster variance, projected dimensionality, projected distance, and projected radius. In terms of experimental results, we compare our scheme with HDDStream for cluster purity, memory usage, and cluster's sensitivity.

\section{Problem Formulation}\label{sec:ProblemFormulation}
In general, data stream is modeled as an infinite series of points $\{\mathbf{p}_{1},\mathbf{p}_{2},...,\mathbf{p}_{i},...\}$ arriving at discrete time $\{t_{1},t_{2},...t_{i},...\}$. Each point $\mathbf{p}_{i}$ is a vector of dimension $d$ such that $\mathbf{p}_{i}=\{p_{i,1},p_{i,2},...,p_{i,d}\}$.

An important characteristic of data streams is that we cannot store all data points. A usual way to overcome this problem is to summarize the data through an appropriate summary structure, often called micro-cluster. A micro-cluster summarizes the time and dimensionality limited stream data in the form of a tuple. When aging is also under consideration, the temporal extension of micro clusters \cite{aggarwal_framework_2004} is employed.
Recent research works \cite{aggarwal_framework_2004}, \cite{ntoutsi_density-based_2012} use the following definition of micro-cluster:\\
\textbf{Definition 0.} (Micro-cluster $mc$)\\
A micro-cluster at time $t$ for a set of d-dimensional data points $\mathcal{C}=\{\mathbf{p}_{0},\mathbf{p}_{1},...,\mathbf{p}_{N-1}\}$ arriving at discrete time $t_{0},t_{1},...,t_{N-1}$, is summarized as ($2d+1$) size tuple $mc(t)=\{\mathbf{CF1}(t),\mathbf{CF2}(t),W(t)\}$, where $\mathbf{CF1}(t)$ and $\mathbf{CF2}(t)$ are $d$ dimensional vectors, defined as:
\begin{itemize}
  \item $\mathbf{CF1}(t)$ is the d-dimensional vector of weighted sum of points $\{\mathbf{p}_{1},\mathbf{p}_{2},...,\mathbf{p}_{i},...\}$ along each dimension, such that for dimension $j$ we have $CF1_{j}=\sum_{i=0}^{N-1}p_{i,j}f(t-t_{i})$, where $N$ is the size of time window, $p_{i,j}$ is the $i^{th}$ point in time window and $f(t-t_{i})$ is the weight of the $i^{th}$ point.
  \item $\mathbf{CF2}(t)$ is the d-dimensional vector of weighted sum of the squares of the points $\{\mathbf{p}_{1},\mathbf{p}_{2},...,\mathbf{p}_{i},...\}$ along each dimension, such that for dimension $j$ we have $CF2_{j}=\sum_{i=0}^{N-1}p_{i,j}^{2}f(t-t_{i})$, where $N$ is the size of time window, $p_{i,j}$ is the $i^{th}$ point in time window and $f(t-t_{i})$ is the weight of the $i^{th}$ point.
  \item $W(t)$ is the sum of the weights of data points, mathematically, $W(t)=\sum_{i=0}^{N-1}f(t-t_{i})$.
\end{itemize}

\begin{figure}[!t]
\centering
\includegraphics[width=4in]{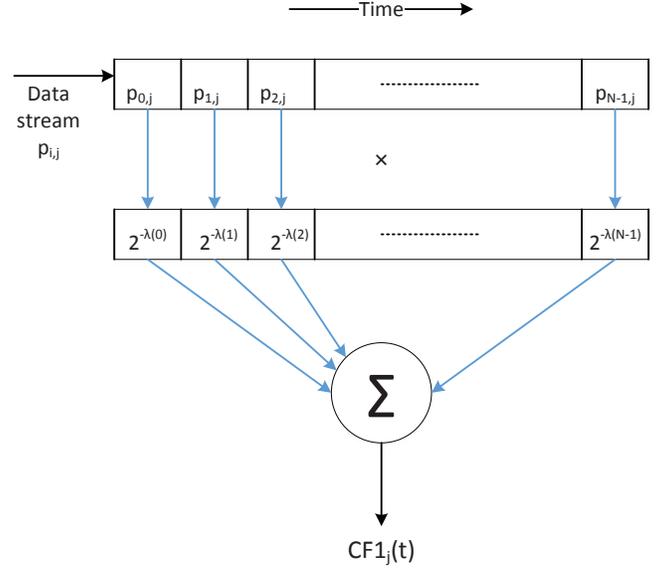}
\caption{Practical approach to update micro-cluster tuple}\label{CAF}
\end{figure}

In data streams, since we are more interested in the data within a certain recent time window instead of all historical data, an  aging effect has been used for weighted function $W(t)$. The recent works \cite{aggarwal_framework_2004}, \cite{ntoutsi_density-based_2012} have used conventional exponential fading function $f(t)=2^{-\lambda t}$, where $\lambda$ is the decay rate. By using fading function $f(t)$ we need to maintain a memory buffer of time window size for each cluster, because, whenever a new point arrives we need to shift the previous data in the buffer of fixed size.
We want to highlight an important point here that the online update of the tuples \cite{aggarwal_framework_2004}, \cite{ntoutsi_density-based_2012} of the form $mc(t)=\{\mathbf{CF1}(t)+p,\mathbf{CF2}(t)+p^{2},W(t)+1\}$ is not practically feasible because it leads to monotonically increasing weighted sum data.  A practical approach for updating the tuple is shown in Fig. \ref{CAF}. It is obvious that for a fixed size memory shift register, when a new point arrives the old point is discarded. The correct mathematical expression for online update, then, becomes, $mc(t)=\{\mathbf{CF1}(t)-\mathbf{CF1}_{N-1}+p,\mathbf{CF2}(t)-\mathbf{CF2}_{N-1}+p^{2},W(t)-f(t-t_{N-1})+1\}$, such that for dimension $j$ we have  $CF1_{N-1,j}=p_{N-1,j}f(t-t_{N-1})$ and $CF2_{N-1,j}=p_{N-1,j}^{2}f(t-t_{N-1})$.

\begin{figure}
  \centering
  \includegraphics[width=3in]{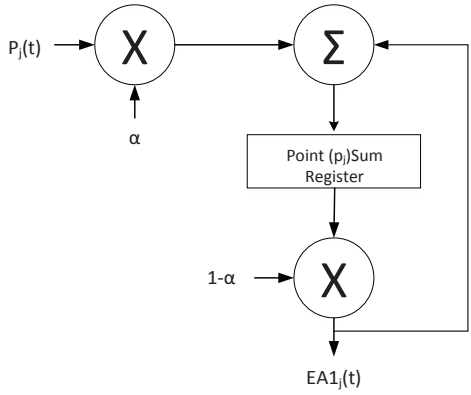}\\
  \caption{Exponential moving average based update of micro-cluster tuple}\label{EA1}
\end{figure}

We define micro-cluster as follows:
\begin{definition} {(Micro-cluster $mc$)}\label{def:mc_EA}
We redefine the micro-cluster as a set of points $\mathcal{C}=\{\mathbf{p}_{0},\mathbf{p}_{1},...,\mathbf{p}_{N-1}\}$ arriving at discrete time points $t_{0},t_{1},...,t_{N-1}$. The $mc$ is summarized as $2d+1$ size tuple $mc(t)=\{\mathbf{EA1}(t),\mathbf{EA2}(t),W(t)\}$, where $\mathbf{EA1}(t)$ and $\mathbf{EA2}(t)$ are $d$ dimensional vectors, defined as:
\begin{itemize}
  \item $\mathbf{EA1}(t)$ is the d-dimensional vector of exponential weighted moving average of points $\{\mathbf{p}_{1},\mathbf{p}_{2},...,\mathbf{p}_{i},...\}$ along each dimension, such that for dimension $j$ we have $EA1_{j}(t)=\alpha p_{j}(t)+(1-\alpha)EA1_{j}(t-1)$, where $\alpha=2/(1+N)$ is a smoothing factor controlled by the size of time window and $p_{j}(t)$ is the latest point in time window.
  \item $\mathbf{EA2}(t)$ is the d-dimensional vector of exponential weighted average of points $\{\mathbf{p}_{1},\mathbf{p}_{2},...,\mathbf{p}_{i},...\}$ along each dimension, such that for dimension $j$ we have $EA2_{j}(t)=\alpha p_{j}^{2}(t)+(1-\alpha)EA2_{j}(t-1)$.
  \item $W(t)$ is the sum of the of data points at time $t$.
\end{itemize}
 \end{definition}
In order to formalize aging effect of data we introduce exponential moving average of data stream within a specified time window. We use exponential weighted moving average in the tuple as decreasing exponential function. Notice that now, calculation of  $\mathbf{EA1}(t)$ or $\mathbf{EA2}(t)$ does not require storage of past values, and only one addition and two multiplications with one memory register (of the size of dimension $j$) are required to update the tuple at any time instance. Design implementation of our micro-cluster update is shown in Fig. \ref{EA1}.\\
Data stream contains high dimensional data where each dimension has its own importance. In order to collate the similar points in data stream we use variance along each dimension. The lower the variance the higher the correlation among the points in particular dimension. We use variance as a metric to limit the number of dimensions to preferred dimensions only.
\begin{definition}{(Preferred Dimension)}
A dimension $j$ is said to be a preferred dimension if $Var_{j}(mc)<\xi$, where $\xi$ is the variance threshold and $Var_{j}(mc)$ is the variance of $mc$
 along dimension $j$, defined as:
 \begin{equation}\label{variance}
   Var_{j}(mc)=EA2_{j}(t)-(EA1_{j}(t))^{2}
 \end{equation}
  \end{definition}
 The preferred dimension helps gather the data points which have preferred dimensions less than a pre-defined threshold. Intuitively, it indicates the similarity across dimensions controlled by the variance threshold ($\xi$). In conjunction with preferred dimension, we define the preferred dimension vector.
\begin{definition}  {(Preferred Dimension Vector)}
 Every micro-cluster has a preferred dimension vector defined as:
 \begin{equation}\label{PDV}
   \Psi(mc)=\{\psi_{1},\psi_{2},...,\psi_{d}\}
 \end{equation}
 with
 \begin{equation}\label{PDVweight}
   \psi_{j}=\left\{
                 \begin{array}{ll}
                   \varrho, & \hbox{$Var_{j}(mc)< \xi$;} \\
                   1, & \hbox{otherwise.}
                 \end{array}
               \right.
 \end{equation}
 where
 $\xi\in \mathbb{R}$, and $\varrho\in \mathbb{R}$ is a constant $\varrho \gg 1$. The number of elements in preferred dimension vector gives the projected dimensionality of the micro-cluster. The term 'projected' differentiates the micro-cluster defined over s projected subspace of the feature space instead of the whole feature space.
 \end{definition}
\begin{definition}  {(Projected Dimensionality)}\label{def:PDIM}
 Let $p\in \mathcal{C}$ and $\xi\in \mathbb{R}$. The number of dimensions $j$ with $Var_{j}(mc)<\xi$ is called projected dimensionality of $mc$ and denoted by P\textsc{DIM}(mc).
\end{definition}
 Weighting the dimensions inversely proportional to their variance is not useful because we are only interested in distinguishing between dimensions with low variance and all other dimensions. Therefore, we use only two-valued weight vector.
 It can be easily determined from the preferred dimension vector by counting the number of dimensions with value $\varrho$. The intuition of calculating projected dimensionality is to find projected core micro-cluster, i.e., the clusters with some subspace of dimensions instead of all dimensions.

\begin{definition} {(Projected Radius)}\label{def:projectedRadius}
  Let $mc$ be a micro-cluster, $\xi\in \mathbb{R}$, and $\varrho\in \mathbb{R}$ is a constant $\varrho \gg 1$. The projected radius of $mc$ is given by:
  \begin{equation}\label{projectedRadius}
    r_{\Psi}(mc)=\sqrt{\sum_{j=1}^{d}\frac{\psi_{j}}{\varrho}\left(EA2_{j}(t)-(EA1_{j}(t))^{2}\right)}
  \end{equation}
  where $\varrho$ normalizes the variance along each dimension. This is the projected radius that takes into account the preferred dimensions of the micro-cluster.
  \end{definition}
\begin{definition} {(Projected Distance)} \label{def:projectedDistance}
Let $p\in \mathcal{D}$ and $mc$ be a projected micro-cluster with dimension preference vector $\Psi(mc)$. The projected distance between $p$ and $mc$ is given by:
\begin{equation}\label{projectedDistance}
  dist^{proj}(p,mc)=\sqrt{\sum_{j=1}^{d}\frac{\psi_{j}}{\xi}(p_{j}-center_{j}^{mc})^{2}}
\end{equation}
where $center^{mc}$ is the center of micro-cluster $mc$ and is given by $center^{mc}=\mathbf{EA1}(t)$.
\end{definition}
  Now we introduce the notion of core-projected $mc$ which is an essential component of density based clustering. A core-projected $mc$ is a $mc$ that contains at least $\mu$ number of points within a projected radius of $\epsilon$ with projected dimensionality less than a threshold $\pi$.
\begin{definition}  {(Core Projected Micro-cluster)} \label{def:coreMc}
Let $\epsilon,\xi\in \mathbb{R}$ and $\pi,\mu\in \mathbb{N}$. A micro-cluster $mc$ is called a core projected $mc$ if the preference dimensionality of $mc$ is at most $\pi$ and it contains at least $\mu$ points within its projected radius $\epsilon$, formally:
\begin{align}\label{coreMc}
\nonumber  & \textsc{Core}^{proj}(mc) \iff \\
  & (r_{\Psi}(mc) <\epsilon) \wedge (W(t) >\mu) \wedge (\textsc{Pdim}<\pi).
\end{align}
In other words, a micro-cluster $mc$ is a core projected $mc$ \emph{iff}:
   \begin{enumerate}[(1)]
     \item $r_{\Psi}(mc) <\epsilon$
     \item $W(t) >\mu$
     \item $\textsc{Pdim}<\pi$
   \end{enumerate}
\end{definition}
There might be micro-clusters that do not fulfill the above constraints either because their associated number of points is smaller than $\mu$ or because their projected dimensionality exceeds $\pi$. These micro-clusters are treated as outliers.
\begin{definition} {(Outlier Micro-cluster)}
Let $\epsilon,\xi\in \mathbb{R}$ and $\pi,\mu\in \mathbb{N}$. A micro-cluster $mc$ is called a outlier $mc$, if its projected dimensionality is at least $\pi$ and its projected radius and $\epsilon$-Neighbors are at most $\epsilon$ and $\mu$, respectively, formally:
\begin{align}\label{outlierMc}
\nonumber  & outlier(mc) \iff \\
& (\textsc{Pdim}>\pi) \wedge (r_{\Psi}(mc)<\epsilon) \wedge (W(t)<\mu).
\end{align}
\end{definition}
In order to keep update the micro-clusters, i.e., to check for possible conversion of core micro-cluster to outlier micro-cluster and vice versa we introduce an outlier threshold ($0<\beta <1$) such that an outlier micro-cluster becomes a potential core micro-cluster if $W>\beta \mu$ in addition to the conditions in \eqref{coreMc} . Similarly, a core micro-cluster becomes a potential outlier micro-cluster if $W< \beta \mu$ in addition to the conditions in \eqref{outlierMc}.
The micro-cluster can be easily maintained online when a new point arrives in a cluster and other $mc$ need time degradation.
\begin{remark}{(Online maintenance)}
The micro-cluster $mc$ defined in \emph{definition 1} holds simple additive property that facilitates the online maintenance.
\begin{itemize}
  \item If a point $p$ arrives at time $t$, then the updated tuple is given by $mc(t)=\{\alpha p+(1-\alpha)EA1(t-1),\alpha p^{2}+(1-\alpha)EA2(t-1),W(t-1)+1\}$.
  \item If no point adds in a micro-cluster at time $t$, then the updated tuple is given by $mc(t)=\{(1-\alpha)EA1(t-1),(1-\alpha)EA2(t-1),W(t-1)\}$.
\end{itemize}
\end{remark}
\section{The HSDStream Algorithm}
HSDStream algorithm can be divided into three parts: 1) initialization to produce a set of representative core micro-cluster (core-mc) from an initial chunk of data points, 2) online maintenance of core-mc and outlier micro-cluster (outlier-mc), and, 3) offline generating the final clusters, on demand by the user.
\subsection{Initialization}
\begin{figure}[!t]
  \centering
  \includegraphics[width=3.5in]{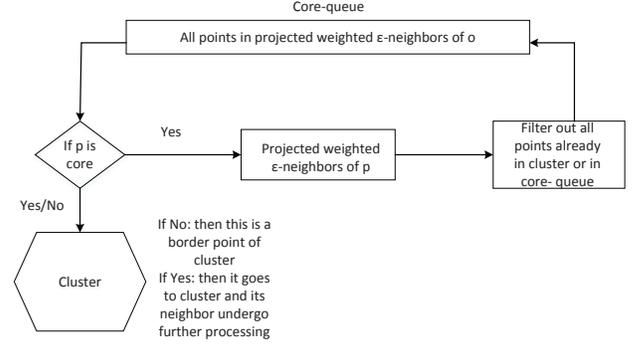}\\
  \caption{Generation of initial set of micro-clusters}\label{initialization}
\end{figure}

In order to get initial set of micro-clusters from a fixed size of data points, we apply density-based projected clustering algorithm, a variant of PreDeCon algorithm \cite{bohm_density_2004}, which is designed to work for fixed size of data of high dimensionality. Let $\mathcal{D}$ be a set of initial chunk of d-dimensional data points ($\mathcal{D}\subseteq \mathbb{R}^{d}$). For each point $p\in \mathcal{D}$, we find a set of $\epsilon-$neighbors $\mathcal{N}_{\epsilon}(p)$. In addition to this, we find the neighbors of $p$ with projected distance equal to or less than the $\epsilon$, namely, $\mathcal{N}_{\epsilon}^{\Psi(p)}(p)$.
\begin{definition} {(Projected Distance of a Point)}
Let $p,q\in \mathcal{D}$. The projected distance of a point $p$ with any point $q$ is given by:
\begin{equation}\label{dist}
  dist_{p}(p,q)=\sqrt{\sum_{i=1}^{d}\frac{\psi_{i}(p)}{\varrho}\left(d_{i}(p)-d_{i}(q)\right)^{2}}
\end{equation}
where $d_{i}(p)$ is the $i^{th}$ dimension of point $p$. Note that, in general $dist_{p}(p,q)\neq dist_{p}(q,p)$ because of the projected dimension vectors of point $p$ and $q$. In order to get symmetrical distance between $p$ and $q$ we use maximum of $dist_{p}(p,q)$ and $dist_{p}(q,p)$.
\end{definition}
A projected core point $o\in \mathcal{D}$ can be defined with the same intuition of projected micro-cluster in definition \ref{def:coreMc}.
\begin{equation}\label{corePoint}
  \textsc{Core}^{proj}(o) \iff \textsc{Pdim}(\mathcal{N}_{\epsilon}(o))\leq \pi \wedge |\mathcal{N}_{\epsilon}^{\Psi(p)}(o)|\geq \mu
\end{equation}
The initialization function in algorithm \ref{algorithm1} line \ref{alg1:initializationFn} runs the algorithm for the creation of initial set of $mc$. It starts  by inserting all points in the set $\mathcal{N}_{\epsilon}(o)$ into a queue. For each point in the queue, it computes all directly projected weighted reachable points and inserts those points into the queue which are still unclassified. This process repeats until the queue is empty and the cluster is computed. The flow chart of algorithm is shown in Fig. \ref{initialization}. Remove all those points belong to calculated cluster from dataset $\mathcal{D}$ and repeat the process for another core point. This process remains continue till all the core points are exhausted.

\subsection{Real-time Maintenance of Micro-clusters}
In order to find out the clusters in an evolving real-time data stream, we maintain two groups of micro-clusters, namely, core-mc and outlier-mc in real-time. All the micro-clusters are maintained in a separate memory space. A new point might be assigned to core-mc, outlier-mc, or it may start new outlier-mc depends upon various factor. Sequential process of merging a new point $p$ is described below:
\begin{enumerate}
  \item When a new point arrives, it first becomes the candidate of core-mc (algorithm \ref{algorithm1}, line \ref{alg1:addpToCoreMc}). The projected dimensionality of each core-mc has been evaluated before and after adding this point $p$ (algorithm \ref{algorithm2}, line \ref{alg2:PDIM}). After that, projected distance of $p$ is calculated with those core-mc which still satisfy the projected dimensionality constraint, i.e., after the addition of point $p$ (algorithm \ref{algorithm2}, line \ref{alg2:projectedDistance}). Then, we choose one core-mc which has smallest projected distance from $p$ (algorithm \ref{algorithm2}, line \ref{alg2:core_mc_closest}). Finally, the projected radius of chosen core-mc ($p$ included) has been evaluated and checked for upper bound $(\epsilon)$ (algorithm \ref{algorithm2}, line \ref{alg2:ifr}). If it satisfies, then point $p$ is assigned to that core-mc (algorithm \ref{algorithm2}, line \ref{alg2:updateTuple_fn} using update tuple function in algorithm \ref{algorithm4}), else it becomes candidate of outlier-mc list.
  \item When a new point becomes a candidate for an outlier-mc, the projected distance of $p$ with each outlier-mc has been evaluated (algorithm \ref{algorithm3}, line \ref{alg3:projectedDistance}). The closest distant outlier-mc is chosen in line \ref{alg3:core_mc_closest}. The point $p$ becomes the member of that outlier-mc if the projected radius is less than or equal to the radius threshold ($\epsilon$) (algorithm \ref{algorithm3}, line \ref{alg3:updateTupleFn}). In order to get long term effect we check the possibility of outlier-mc to core-mc conversion after certain number of points (window size $N$).
  \item If point $p$ cannot be added in core-mc or outlier-mc (algorithm \ref{algorithm3}, line \ref{alg3:trial_outlier0}) then a new outlier-mc is created with this point being the first element. It may become the seed of future core-mc.
\end{enumerate}
\begin{algorithm}
\footnotesize
\caption{HSDStream main}
\label{algorithm1}
\begin{algorithmic}[1]
\STATE  \emph{Initialization}
\STATE initial parameters $\pi,\xi,\epsilon,N$
\STATE $data stream=\{\mathbf{p}_{1},\mathbf{p}_{2},...,\mathbf{p}_{i},...\}$
\STATE $initialBuffer=readData(numOfIntialPoints)$
\STATE $core\_mc=initialization\_fn(initialBuffer)$ \label{alg1:initializationFn}
\FOR{$i=1$ to $numOfMc$}
\STATE $mcTuple=createMcTuples(core\_mc)$ \\
\COMMENT{It creates $mcTuple=\{\mathbf{EA1}(t),\mathbf{EA2}(t),W(t)\}$, an $numOfMc\times (2d+1)$ matrix}
\ENDFOR
\WHILE{Stream has data points}
\STATE $windowBuffer=readData(N)$ 
\FOR{$i=1$ to $N$}
\STATE $p_{i}=windowBuffer(i)$  // i-th point from windowBuffer
\STATE $[trial\_core,mcTuple]=addpToCoreMc(p_{i},mcTuple)$  \label{alg1:addpToCoreMc}
\IF{$trial\_core==1$}
\STATE Degrade all outlierTuples
\ELSE
\STATE $[trial\_outlier,outlierTuple]=addpToOutlierMc(p_{i},outlierTuple)$
\ENDIF
\IF{$trial\_core==0 \; \&\& \; trial\_outlier==0$}
\STATE $newOutlierMc=createOutlierMc(p_{i})$
\STATE update outlierTuple list
\ENDIF
\ENDFOR \\
\COMMENT{core-mc to outlier-mc conversion}
\STATE $[movedMcTuples,remainingMcTuples]=moveMcTuples(mcTuples)$ \\
\COMMENT{outlier-mc to core-mc conversion}
\STATE $[movedOutlierTuples,remainingOutlierTuples]=moveOutlierTuples(outlierTuples)$
\STATE $updatedMcTuples=remainingMcTuples+movedOutlierTuples$
\STATE $updatedOutlierTuples=remainingOutlierTuples+movedMcTuples$
\ENDWHILE
\end{algorithmic}
\end{algorithm}

\begin{algorithm}
\footnotesize
\caption{Add data point to core-mc}
\label{algorithm2}
\begin{algorithmic}[1]
\STATE  \emph{addpToCoreMc(p,mcTuples)}
\FOR{$i=1$ to $numOfTuples$}
\STATE $updatedTuples=updateTuple\_fn(p,mcTuple(i))$
\STATE Calculate updated \textsc{Pdim} // using definition \ref{def:PDIM} \label{alg2:PDIM}
\IF{$\textsc{Pdim}\leq \pi$}
\STATE Calculate projected distance // using definition \ref{def:projectedDistance} \label{alg2:projectedDistance}
\ENDIF
\ENDFOR
\STATE $core\_mc\_closest=min(projectedDistances)$ \label{alg2:core_mc_closest}
\STATE Calculate projected radius $r_{\Psi}(core\_mc\_closest)$ // using definition \ref{def:projectedRadius}
\IF{$r_{\Psi}(core\_mc\_closest)< \epsilon$} \label{alg2:ifr}
\STATE $mcTuple=updateTuple\_fn(p,mcTuple)$  \label{alg2:updateTuple_fn}
\STATE Update all other mcTuples with one degradation
\RETURN{$trial\_core=1$}
\ELSE
\STATE Degrade all mcTuples
\RETURN{$trial\_core=0$}
\ENDIF
\end{algorithmic}
\end{algorithm}

\begin{algorithm}
\footnotesize
\caption{Add data point to outlier-mc}
\label{algorithm3}
\begin{algorithmic}[1]
\STATE  \emph{addpToCoreMc(p,oulierTuples)}
\FOR{$i=1$ to $numOfOutlierTuples$}
\STATE $updatedTuples=updateTuple\_fn(p,mcTuple(i))$
\STATE Calculate projected distance // using definition \ref{def:projectedDistance} \label{alg3:projectedDistance}
\ENDFOR
\STATE $core\_mc\_closest=min(projectedDistances)$ \label{alg3:core_mc_closest}
\STATE Calculate projected radius $r_{\Psi}(outlier\_mc\_closest)$ // using definition \ref{def:projectedRadius}
\IF{$r_{\Psi}(outlier\_mc\_closest)< \epsilon$}
\STATE $outlierTuple=updateTuple\_fn(p,outlierTuple)$ \label{alg3:updateTupleFn}
\STATE Update all other outlierTuples with one degradation
\RETURN{$trial\_outlier=1$}
\ELSE
\STATE Degrade all outlierTuples
\RETURN{$trial\_outlier=0$} \label{alg3:trial_outlier0}
\ENDIF
\end{algorithmic}
\end{algorithm}

\begin{algorithm}
\footnotesize
\caption{Update Tuple function}
\label{algorithm4}
\begin{algorithmic}[1]
\STATE  $updateTuple\_fn(p,Tuple)$
\STATE $\mathbf{EA1}(t-1)=Tuple(1:d)$
\STATE $\mathbf{EA2}(t-1)=Tuple(d+1:2d)$
\STATE $W(t-1)=Tuple(end)$
\STATE $\mathbf{EA1}(t)=\alpha \mathbf{p}+(1-\alpha)\mathbf{EA1}(t-1)$
\STATE $\mathbf{EA2}(t)=\alpha \mathbf{p}^{2}+(1-\alpha)\mathbf{EA2}(t-1)$
\STATE $W(t)=W(t-1)+1$
\STATE $newTuple=\{\mathbf{EA1}(t),\mathbf{EA2}(t),W(t)\}$
\end{algorithmic}
\end{algorithm}
\subsection{Clusters Generation: Offline}
The real-time maintained micro-clusters capture the density area and the projected dimensionality of data streams. However, in order to get meaningful clusters, we need to apply some clustering algorithm to get the final result. When a clustering request arrives, a variant of PreDeCon algorithm \cite{bohm_density_2004} is applied on the set of real-time maintained  core-mc(s) to get the final result of clustering. In density-based PreDeCon, a core point starts a micro-cluster, all the directly connected points and the chain of core points which satisfy $\epsilon-$neighborhood criteria and maximum dimensionality $\pi$ become the member of that cluster. During offline on-demand clustering phase, each core-mc acts as core point. Each core-mc is regarded as a virtual point located at the center of core-mc. We use the concept of density connectivity to determine the final clusters. That is, all the density-connected core-mc(s) form a cluster.

\section{Discussion}
In this section we highlight issues and challenges in the development of high dimensional data stream clustering in Internet traffic monitoring. We maintain the density with $\epsilon-$neighborhood and minimum number of points $\mu$ in a core-mc.
When an identical burst of data (in case of attack on network) arrives, outlier-mc(s) are diminished and only one core-mc remains there. In this case, an important entity of core-mc formation i.e., projected dimensionality cannot work because, now $\textsc{Pdim}=d$ and it no longer satisfies the condition $\textsc{Pdim}\leq \pi$. In order to overcome this problem we introduce another condition ORed with the condition $\textsc{Pdim}\leq \pi$ to maintain one core-mc containing exactly similar data. The new condition is $W(t)/N >90\%$ , i.e., if the data points window contains more than $90\%$ points, then no need to check $\textsc{Pdim}$ because the majority of identical data points indicates some abnormal activity on the network being monitored.
During real-time maintenance, when a new point arrives and it becomes a part of only one micro-cluster, then, all the other micro-clusters undergo one time degradation. For each existing core-mc, if no new point is merged into it, then the weight of core-mc will decay gradually. If the weight is below $\beta \mu$, then it means that core-mc has become an outlier-mc, it should be deleted and its memory space should be released for new core-mc. Similarly, if the weight is above $\beta \mu$ then it means that the outlier-mc has become a core-mc, it should be deleted and its memory space should be released. Therefore, we need to check the weight of each micro-cluster periodically. We use a fixed time period to perform this check at every time window interval ($N$). In this way any outlier-mc automatically vanishes if no point merges in it during $N$ time units.

\section{Experimental Evaluation}
We compare our proposed HSDStream algorithm with HDDStream \cite{ntoutsi_density-based_2012} which is the recent projected clustering algorithm for high dimensional data streams. We use corrected KDD 1999 \cite{_kdd-cup-1999-computer-network-intrusion-detection_????} Computer Network Intrusion detection dataset which is typically used for the evaluation of stream clustering algorithms. Both algorithms are implemented in MATLAB and run on Intel i5 Dual Core 2.0GHz with 2 GB RAM.
\subsection{Dataset}
To evaluate the performance of clustering algorithm we use KDD 1999 Network Intrusion detection dataset. This is the dataset used for The Third International Knowledge Discovery and Data Mining Tools Competition, which was held in conjunction with KDD-99 The Fifth International Conference on Knowledge Discovery and Data Mining. It has been reported that original dataset contains bugs, therefore, we use the corrected dataset available online at \url{http://kdd.ics.uci.edu/databases/kddcup99/kddcup99.html}. KDD-CUP'99 Network Intrusion Detection stream dataset which has been used earlier \cite{aggarwal_framework_2003}, \cite{aggarwal_framework_2004}, \cite{cao_density-based_2006}, \cite{ntoutsi_density-based_2012} to evaluate CluSTREAM, HPStream, DenStream, HDDStream, respectively.
This dataset corresponds to the important problem of automatic and real-time detection of network attacks and consists of a series of TCP connection records from two weeks of LAN network traffic managed by MIT Lincoln Labs. Each record can either corresponds to a normal connection, or an intrusion. Most of the connections in this data set are normal, but occasionally there could be a burst of attacks at certain times.
In this dataset, attacks fall into four main categories:
\begin{itemize}
  \item DOS: denial-of-service e.g., syn flood
  \item R2L: unauthorized access from a remote machine, e.g., guessing password
  \item U2R:  unauthorized access to local superuser (root) privileges, e.g., various ``buffer overflow'' attacks
  \item Probing: surveillance and other probing, e.g., port scanning
\end{itemize}
    The attack-types are further classified into one of 24 types, such as back, buffer\_overflow, ftp\_write, guess\_passwd, imap, ipsweep, spy, and so on. It is obvious that each specific attack type can be treated as a sub-cluster. Also, this data set contains totally 494020 connection records, and each connection record has 42 attributes or dimensions that belongs to one of the continuous (35) or symbolic type (7). In the performance analysis of proposed algorithm we use all 35 continuous attributes.
\begin{table}[!ht]
  \centering
  \caption{Parameter values}\label{tab:parameters}
\begin{tabular}{c c}
  \hline
  Parameter & Value\\
  \thickhline
  $N$ & 200 \\
  $\pi$ & 30 \\
  $\mu$ & 10 \\
  $\beta$ & 0.2 \\
  $\xi$ & 0.002 \\
  $initialPoints$ & 1000 \\
  $\epsilon$ & 10\\
  $H$ & 1\\
  \hline
\end{tabular}
\end{table}
\subsection{Cluster Quality Evaluation}
\begin{figure}[!t]
  \centering
  \includegraphics[width=3.5in]{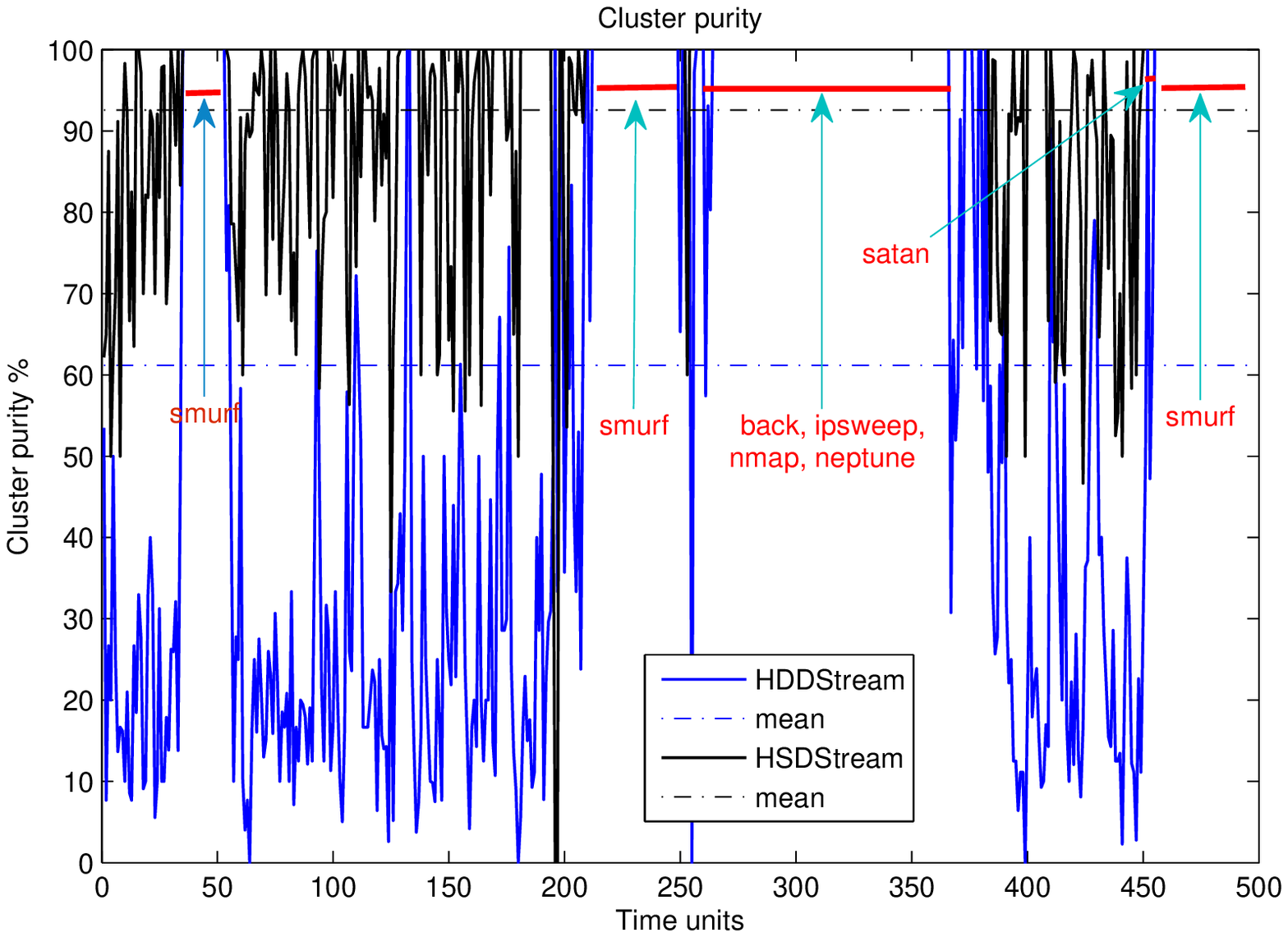}\\
  \caption{Cluster purity with default values of parameters}\label{clusterPurity200Nplot}
\end{figure}

\begin{figure}[!t]
  \centering
  \includegraphics[width=3.5in]{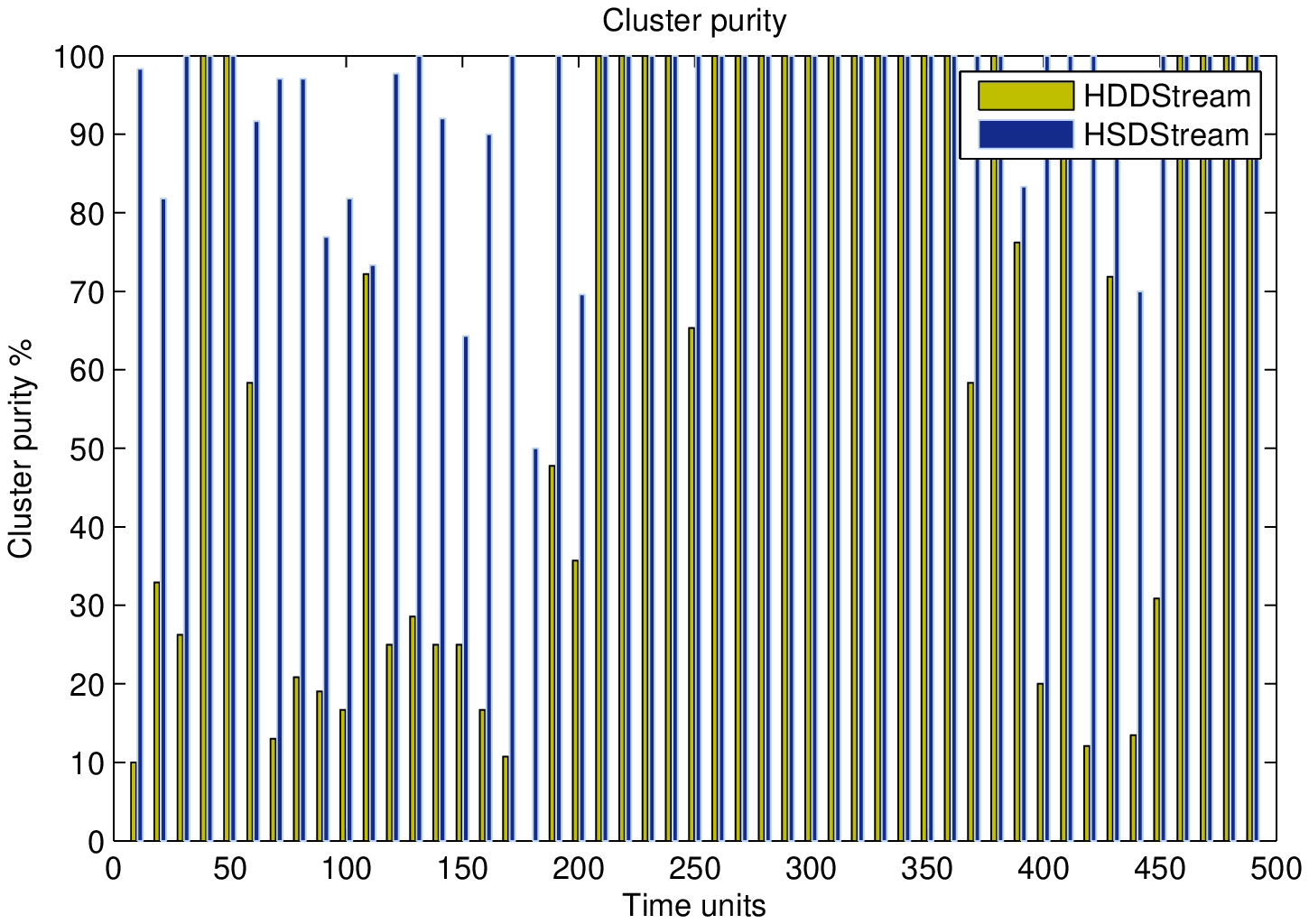}\\
  \caption{Cluster purity with default values of parameters}\label{clusterPurity200Nbar}
\end{figure}

\begin{figure}[!t]
  \centering
  \includegraphics[width=3.5in]{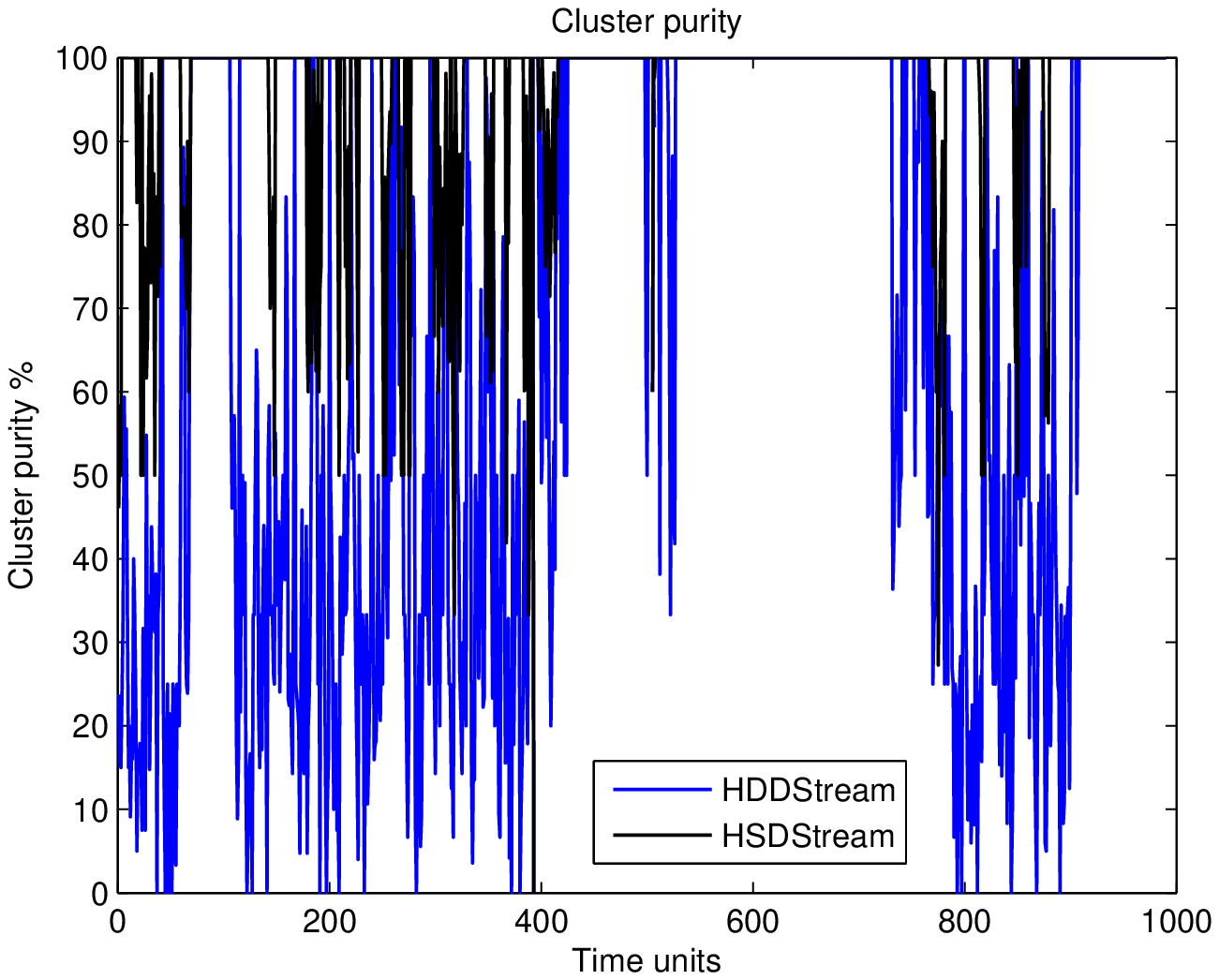}\\
  \caption{Cluster purity with $N=100$}\label{clusterPurity100Nplot}
\end{figure}

\begin{figure}[!t]
  \centering
  \includegraphics[width=3.5in]{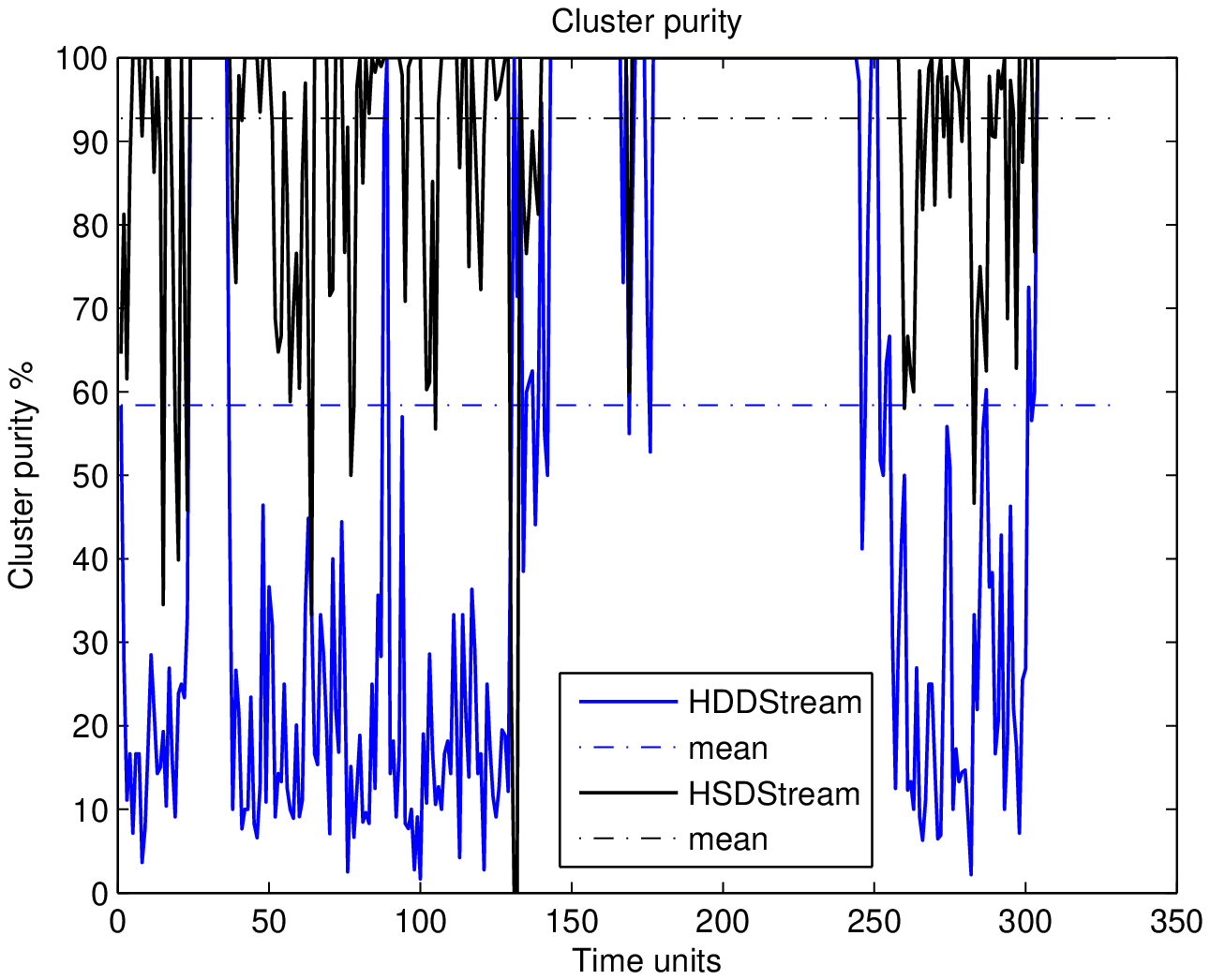}\\
  \caption{Cluster purity with $N=300$}\label{clusterPurity300Nplot}
\end{figure}

\begin{figure}[!t]
  \centering
  \includegraphics[width=3.5in]{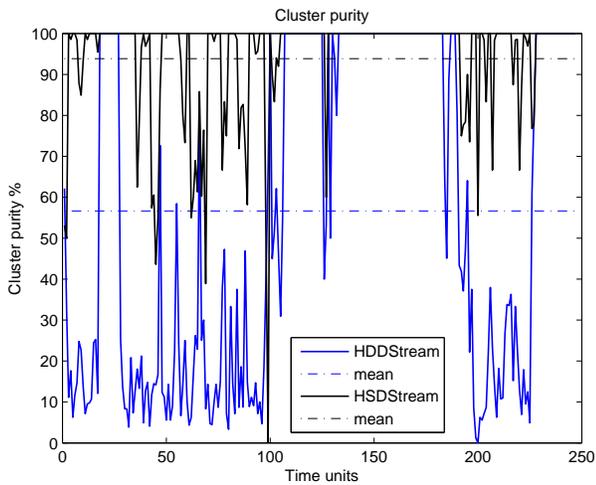}\\
  \caption{Cluster purity with $N=400$}\label{clusterPurity400Nplot}
\end{figure}

Traditional full dimensional clustering algorithms, for example, \cite{aggarwal_framework_2003} used the sum of square distances (SSQ) to evaluate the clustering quality. However, “SSQ is not a good measure in evaluating projected clustering” \cite{aggarwal_framework_2004} because it is a full dimensional measure, and full dimensional measures are not very useful for measuring the quality of a projected clustering algorithm.
So, as in \cite{aggarwal_framework_2004} and \cite{ntoutsi_density-based_2012}, we evaluate the clustering quality by the average purity of clusters, which examines the purity of the clusters with respect to the true cluster (class) labels.
The purity is defined as the average percentage of the dominant class label in each cluster \cite{cao_density-based_2006}. Let there are $K$ number of cluster in a cluster set $\mathcal{K}$ at query time such that $k\in \mathcal{K}=\{1,2,...,K\}$.
\begin{equation}\label{purity}
  purity(\mathcal{K})=\frac{\sum_{k=1}^{K}\frac{|P_{k}^{d}|}{|P_{k}|}}{K}
\end{equation}
where $|P_{k}^{d}|$ is the number of points with dominant class label in cluster $k$ and $|P_{k}|$ is the number of points in cluster $k$. Intuition behind the cluster purity is to measure the actual capture of distinct groups of data points which are known to the given dataset. The time span in which we measure the purity is called \emph{Horizon} window $H$. It is measured in the number of time windows $N$. In the performance analysis $H=1$ otherwise stated.\\
Fig. \ref{clusterPurity200Nplot}-\ref{clusterPurity400Nplot} show the cluster purity of HDDStream and HSDStream. In network streaming data, normal traffic packets (or points) are random in nature at any particular time interval, however, a network attack is characterized by bursts of correlated data packets. Therefore, we cannot fit normal traffic packets in a single cluster. We can fine tune the design parameters ($\alpha$, $\beta$, $\xi$, $N$) to capture the known types of attacks or even the unknown abnormal traffic patterns. We can see that cluster purity can take values from 0 to 1. Cluster purity for normal network traffic usually varies from 0.5 to 1. It can go below 0.5 if we have more than $50\%$ data points with more than $20\%$ dimensions outside the standard deviation of cluster in a certain time window. Intuitively, cluster purity is low if the cluster contains uncorrelated data or in other words, the normal data traffic. High purity (or purity 1) corresponds to highly correlated data as a result of some network attack. In  Fig. \ref{clusterPurity200Nplot}, \emph{smurf} attack can be seen between $34-57$ time units (for $N=200$) which corresponds to data points $7795$ to $11489$ in the KDD network intrusion database. The network is again under \emph{smurf} attack from $211$ to $249$ time units. During the time interval from $250$ to $365$ we encounter with several attacks (\emph{back}, \emph{ipsweep}, \emph{nmap}, and \emph{neptune}) along with correlated normal data so that we can see cluster purity is equal to 1 for this time interval. \emph{Satan} attacks the network from $453$ to $455$ time units, followed by \emph{smurf} attack which continues till the end of simulations at $495$ time units. It can be observed that HDDStream has the same purity graph pattern as HSDStream but with considerably low magnitude. This is due to the large number of core-mc(s) in HDDStream and the fact that percentage purity is inversely proportional to the number of clusters \eqref{purity}. The average cluster purity for HSDStream is $92.57\%$ as compared to the $61.18\%$ of HDDStream.
Next we illustrate: Why HSDStream has fewer number of clusters compared to HDDStream. Since the velocity of points is same for both schemes, it implies that HSDStream has more points per cluster than the HDDStream. For HSDStream, mean value of points in a window $N$ is given by
\begin{align}\label{generalMeanHSD}
 \mathbf{EA1}(n)=&\alpha (1-\alpha)\mathbf{p}_{n}^{n-n}+\alpha (1-\alpha)^{n-\overline{n-1}}\mathbf{p}_{n-1}\\
 \nonumber & +\alpha (1-\alpha)^{n-\overline{n-2}}\mathbf{p}_{n-2}+\ldots +\alpha(1-\alpha)^{n}\mathbf{p}_{0}
\end{align}
where $\alpha=2/(1+N)$. Let $N=200$ and $n=\{0,1,2,...199\}$ with $0$ being the first point and $199$ is the latest point in a buffer window.
Similarly, the mean value of points in window $N$ is given by
\begin{align}\label{generalMeanHDD}
\nonumber  \frac{\mathbf{CF1}(n)}{W}=&\frac{2^{-\lambda (\frac{n-n}{N})}}{W}\mathbf{p}_{n}+\frac{2^{-\lambda   (\frac{n-\overline{n-1}}{N})}}{W}\mathbf{p}_{n-1}\\
\nonumber  &+\frac{2^{-\lambda (n-\overline{n-2})}}{W}\mathbf{p}_{n-2}
  +\ldots +\frac{2^{-\lambda (\frac{n}{N})}}{W}p_{0}
\end{align}
Substituting the values of parameters, we get $\mathbf{EA1}(199)=0.01p_{199}+0.009p_{198}+0.0098p_{197}+...+0.0014p_{0}$ and $\mathbf{CF1}/W=0.0054p_{199}+0.0054p_{198}+0.0054p_{197}+...+0.0046p_{0}$. Thus, for the same point HSDStream gives larger mean value than HDDStream. From equation \eqref{projectedDistance}, it is obvious that higher values of mean (center) result in smaller projected distance, hence larger number of points per cluster and fewer number of clusters.$\blacksquare$ \\
Fig. \ref{clusterPurity200Nbar} depicts the cluster purity for default values of parameters in bar graph. It can be noticed that HSDStream and HDDStream are equally good in detecting the attacked points but the cluster purity for normal traffic is low in HDDStream because of large number of clusters (low density clusters).
Fig. \ref{clusterPurity100Nplot} shows the cluster purity with $N=100$. By decreasing the window size we actually increase the granularity and can capture smaller attacks. The price for this granularity is the more processing for the same amount of data. Again the average value of cluster purity for HSDStream is significantly larger than the HDDStream: $95.23$ versus $67.31$. Fig. \ref{clusterPurity300Nplot} and Fig. \ref{clusterPurity400Nplot}  show the cluster purities for $N=300$ and $N=400$, respectively. We notice that the changing window size has minimal effect on the average cluster purity.



\subsection{Memory Usage}
\begin{figure}[!t]
  \centering
  \includegraphics[width=3.5in]{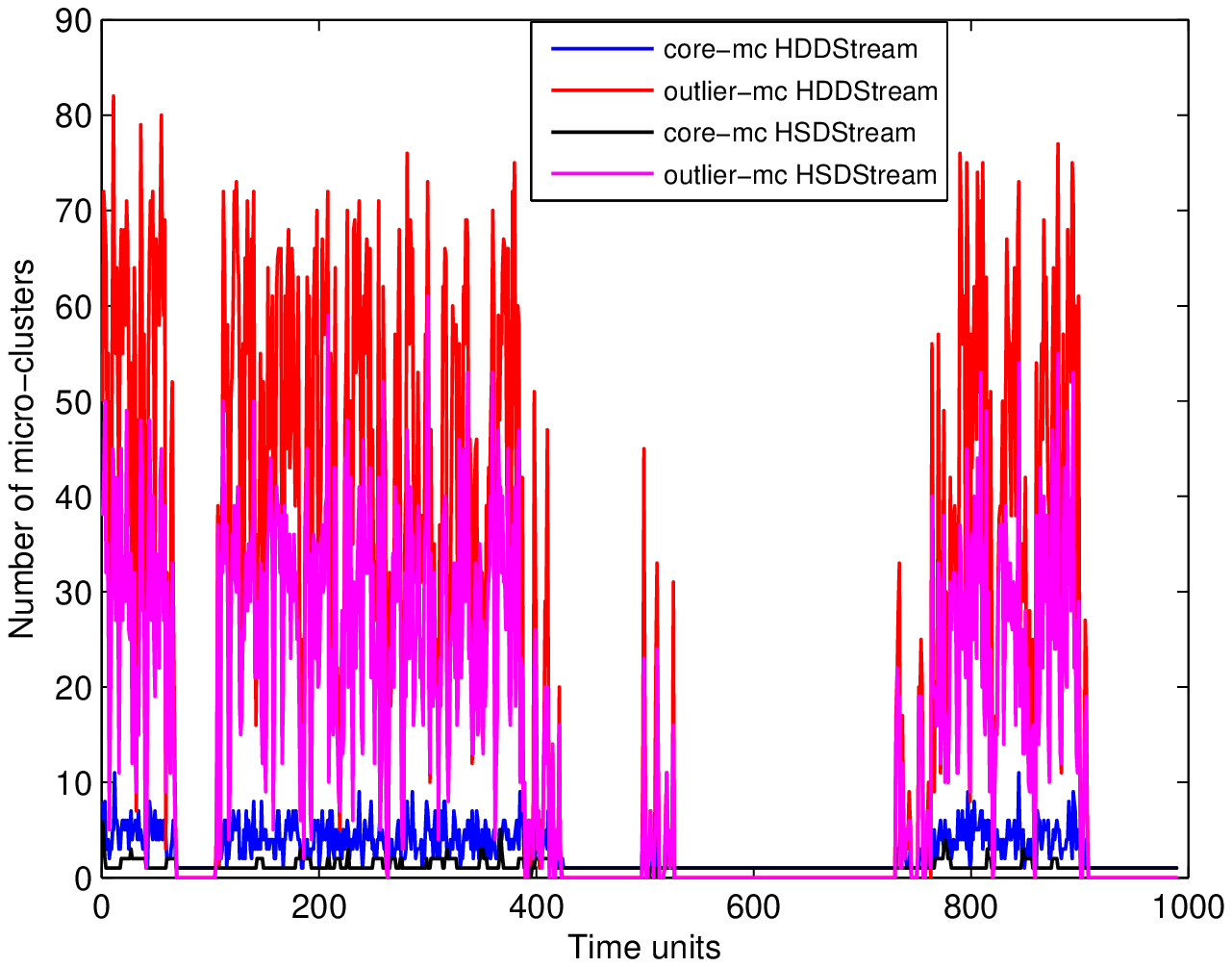}\\
  \caption{Number of clusters with $N=100$}\label{numOfClusters100N}
\end{figure}

\begin{figure}[!t]
  \centering
  \includegraphics[width=3.5in]{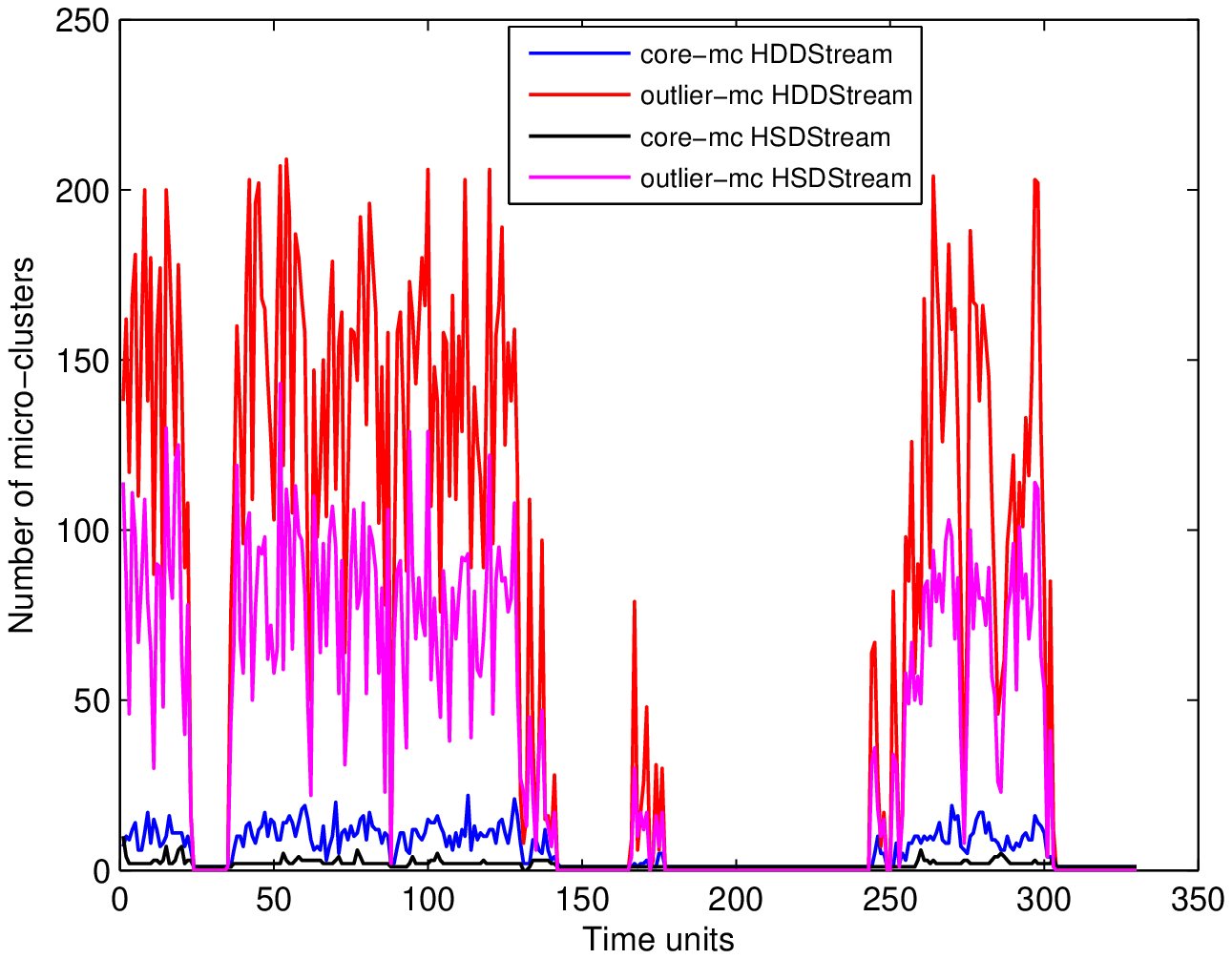}\\
  \caption{Number of clusters with $N=300$}\label{numOfClusters300N}
\end{figure}

\begin{figure}[!t]
  \centering
  \includegraphics[width=3.5in]{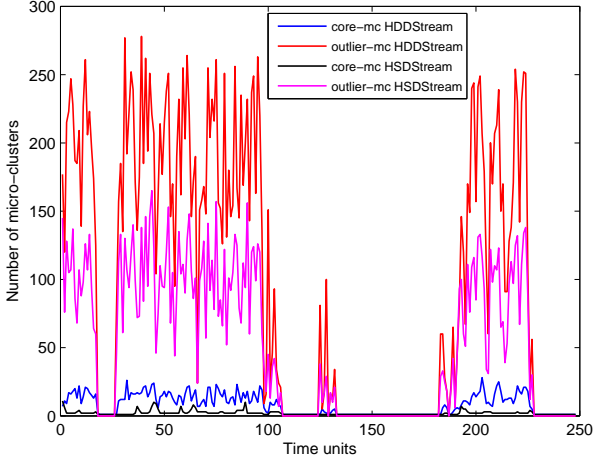}\\
  \caption{Number of clusters with $N=400$}\label{numOfClusters400N}
\end{figure}

We measure the memory usage as a number of micro-clusters in HDDStream and HDSStream. During the period of highly correlated normal data or the network attack, there is only one core-mc containing all the correlated points and no outlier cluster exists. It can be seen from the Figs. \ref{numOfClusters200N}, \ref{numOfClusters100N}, \ref{numOfClusters300N}, and \ref{numOfClusters400N} that the total number of clusters is reduced to one during network attacks. When we compare these figures with different window sizes, we can see that there is a gradual increase of number of clusters with increasing number of window size. HSDStream outperforms the HDDStream in terms of memory usage for all window sizes, which is due to our reduced memory sized tuple and high density micro-clusters. Theoretically, the online update of $CF1_{j}$ requires $N$ number of memory registers (one for each point's $j^{th}$ dimension), whereas, $EA1_{j}$ needs only one memory register, as shown in Fig. \ref{CAF} and Fig. \ref{EA1}, respectively.
\begin{figure}[!t]
  \centering
  \includegraphics[width=3.5in]{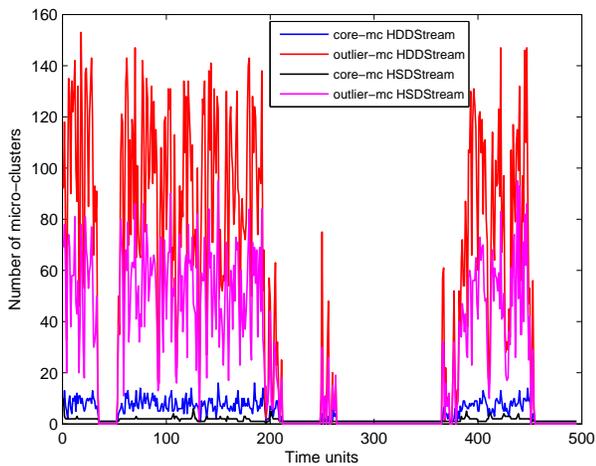}\\
  \caption{Number of clusters with default values of parameters}\label{numOfClusters200N}
\end{figure}
\begin{figure}[!t]
  \centering
  \includegraphics[width=3.5in]{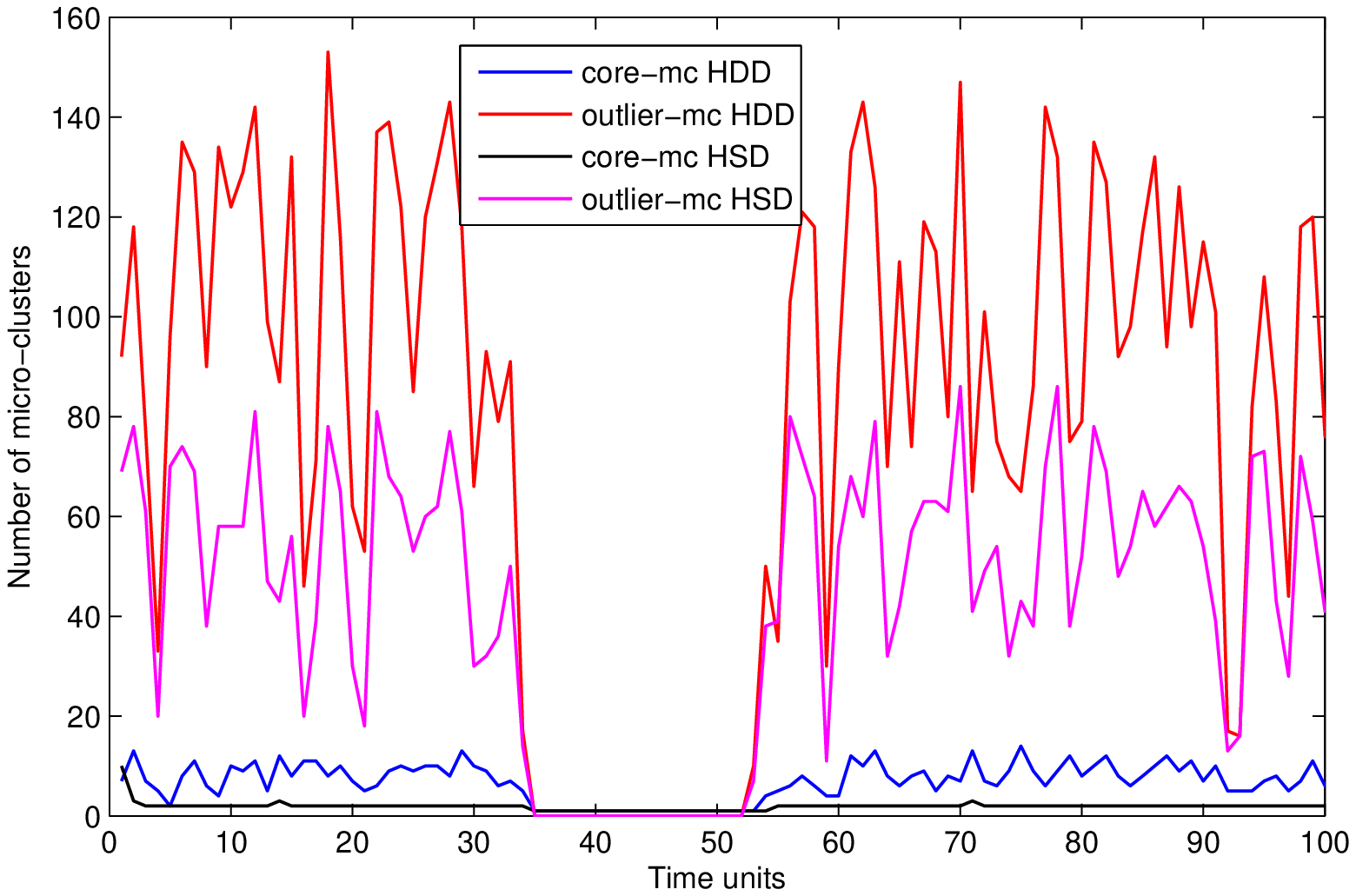}\\
  \caption{Number of clusters with default values of parameters with zoom in}\label{numOfClusters200N}
\end{figure}

\subsection{Sensitivity and Delay Analysis}
\begin{figure}
  \centering
  \includegraphics[width=3.5in]{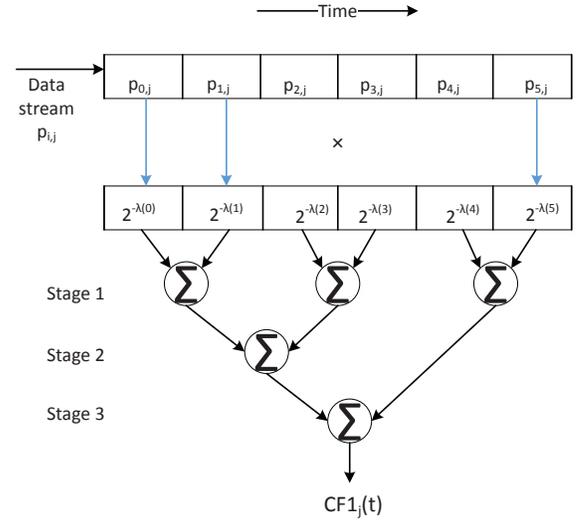}\\
  \caption{Processing time delay in conventional micro-cluster update with $N=6$}\label{CF1delay}
\end{figure}

\begin{figure}[!t]
  \centering
  \includegraphics[width=3.5in]{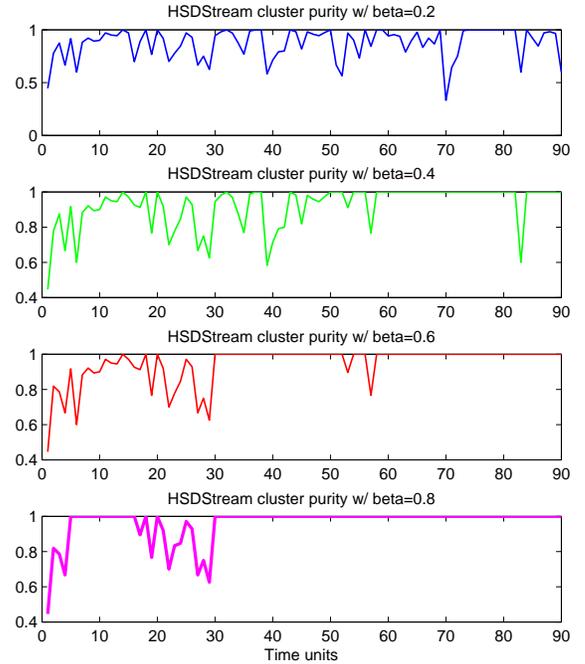}\\
  \caption{Cluster purity for different values of $\beta$}\label{clusterPurity200NwithBetaplot}
\end{figure}

\begin{figure}[!t]
  \centering
  \includegraphics[width=3.5in]{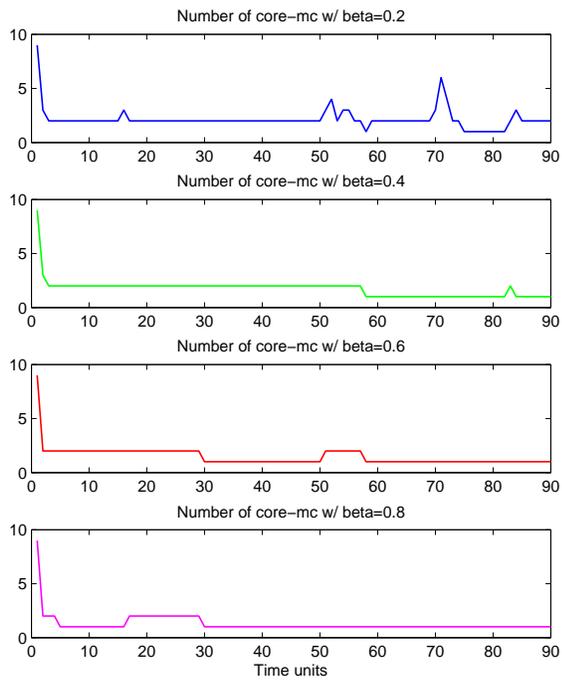}\\
  \caption{Number of core-mc in HSDStream versus $\beta$}\label{numOfMcHSD200NwithBeta}
\end{figure}

\begin{figure}[!t]
  \centering
  \includegraphics[width=3.5in]{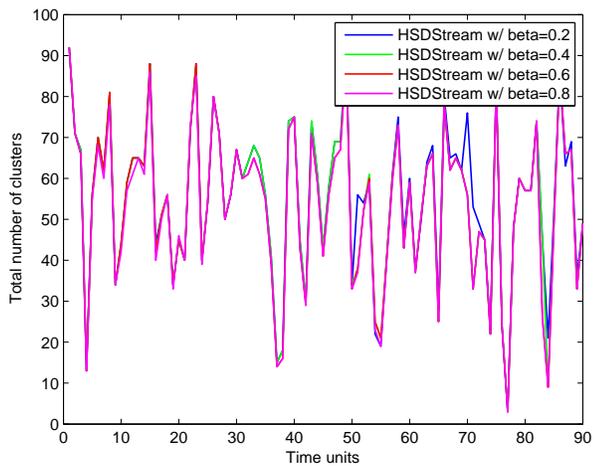}\\
  \caption{Total number of clusters in HSDStream versus $\beta$}\label{numOfClustersHSD200NwithBeta}
\end{figure}

\begin{figure}[!t]
  \centering
  \includegraphics[width=3.5in]{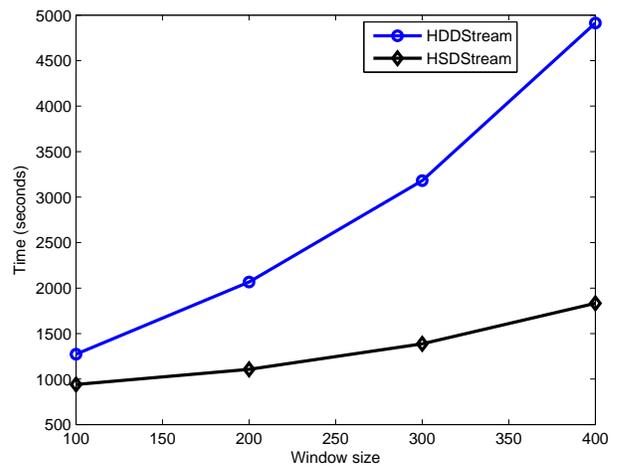}\\
  \caption{Processing time for different window sizes}\label{delay}
\end{figure}

In sensitivity analysis, we show how sensitive the clustering quality is in relevance to the outlier threshold $\beta$, and the processing time with different window sizes. In Fig. \ref{clusterPurity200NwithBetaplot} we see that cluster purity improves with increasing values of outlier threshold. Outlier threshold controls the limit of the number of points that make it eligible to become core-mc or outlier-mc.  After the end of each window size, all micro-clusters are examine for their eligibility as core or outlier. For small values of $\beta$, a cluster remains its current state for the larger time duration making cluster pollute for larger duration. Whereas with high values of $\beta$ the cluster changes its state more quickly (as soon as it violate the condition $NumOfPoints> or <\beta \mu$) leaving the cluster more pure. Fig. \ref{numOfMcHSD200NwithBeta} shows an important result that we can decease the memory usage by increasing the outlier threshold. Higher values of $\beta$ help remove the outlier points thus reducing the unnecessary core-mc(s). Since the core-mc(s) are small proportion of total number of clusters as shown in Fig. \ref{numOfClusters200N}, therefore, the total number of clusters do not exhibit significant improvement in Fig. \ref{numOfClustersHSD200NwithBeta}. However, the memory usage argument remains still valid because core-mc(s) are highly dense and utilize large proportion of memory.\\
Finally, we examine the processing time of HDDStream and HSDStream for different window sizes in Fig. \ref{delay}. This processing time includes the time for the initialization phase and the data collection for the plotting purpose. It can be seen that HSDStream outperforms the HDDStream for all window sizes. This verifies the efficiency of our micro-cluster design in definition \ref{def:mc_EA} where we need only two multipliers and one adder as compared to the conventional micro-cluster defined in definition $0$ which requires $N$ number of multipliers and $N-1$ number of adders with $\lfloor \text{log}_{2}(N)\rfloor$ stages delay. For example if $N=6$, then in order to add $6$ numbers, we need $5$ adders which incur $3$ stages delay as shown in Fig. \ref{CF1delay}.


\FloatBarrier 
\section{Conclusion}
This paper presents a clustering algorithm for high dimensional high density streaming data. We propose a new structure of micro-cluster's tuples. This structure uses exponential weighted averages to reduce the memory usage and decrease the computational complexity. We have compared our scheme with HDDStream with KDD network intrusion detection dataset. The results show that HSDStream give significant improvement over HDDStream in terms of cluster purity, memory usage, and the processing time.



\ifCLASSOPTIONcaptionsoff
  \newpage
\fi


\bibliographystyle{IEEEtran}
\bibliography{sdmbib}

\begin{thebibliography}{10}
\providecommand{\url}[1]{#1}
\csname url@samestyle\endcsname
\providecommand{\newblock}{\relax}
\providecommand{\bibinfo}[2]{#2}
\providecommand{\BIBentrySTDinterwordspacing}{\spaceskip=0pt\relax}
\providecommand{\BIBentryALTinterwordstretchfactor}{4}
\providecommand{\BIBentryALTinterwordspacing}{\spaceskip=\fontdimen2\font plus
\BIBentryALTinterwordstretchfactor\fontdimen3\font minus
  \fontdimen4\font\relax}
\providecommand{\BIBforeignlanguage}[2]{{%
\expandafter\ifx\csname l@#1\endcsname\relax
\typeout{** WARNING: IEEEtran.bst: No hyphenation pattern has been}%
\typeout{** loaded for the language `#1'. Using the pattern for}%
\typeout{** the default language instead.}%
\else
\language=\csname l@#1\endcsname
\fi
#2}}
\providecommand{\BIBdecl}{\relax}
\BIBdecl

\bibitem{forestiero_single_2013}
\BIBentryALTinterwordspacing
A.~Forestiero, C.~Pizzuti, and G.~Spezzano, ``\BIBforeignlanguage{en}{A single
  pass algorithm for clustering evolving data streams based on swarm
  intelligence},'' \emph{\BIBforeignlanguage{en}{Data Mining and Knowledge
  Discovery}}, vol.~26, no.~1, pp. 1--26, Jan. 2013. [Online]. Available:
  \url{http://link.springer.com/10.1007/s10618-011-0242-x}
\BIBentrySTDinterwordspacing

\bibitem{sim_survey_2013}
\BIBentryALTinterwordspacing
K.~Sim, V.~Gopalkrishnan, A.~Zimek, and G.~Cong, ``\BIBforeignlanguage{en}{A
  survey on enhanced subspace clustering},'' \emph{\BIBforeignlanguage{en}{Data
  Mining and Knowledge Discovery}}, vol.~26, no.~2, pp. 332--397, Mar. 2013.
  [Online]. Available: \url{http://link.springer.com/10.1007/s10618-012-0258-x}
\BIBentrySTDinterwordspacing

\bibitem{aggarwal_segment-based_2012}
\BIBentryALTinterwordspacing
C.~C. Aggarwal, ``\BIBforeignlanguage{en}{A segment-based framework for
  modeling and mining data streams},'' \emph{\BIBforeignlanguage{en}{Knowledge
  and Information Systems}}, vol.~30, no.~1, pp. 1--29, Jan. 2012. [Online].
  Available: \url{http://link.springer.com/10.1007/s10115-010-0366-0}
\BIBentrySTDinterwordspacing

\bibitem{amini_density-based_2014}
\BIBentryALTinterwordspacing
A.~Amini, T.~Y. Wah, and H.~Saboohi, ``On density-based data streams clustering
  algorithms: A survey,'' \emph{Journal of Computer Science and Technology},
  vol.~29, no.~1, pp. 116--141, 2014. [Online]. Available:
  \url{http://link.springer.com/article/10.1007/s11390-014-1416-y}
\BIBentrySTDinterwordspacing

\bibitem{jain_data_2010}
\BIBentryALTinterwordspacing
A.~K. Jain, ``\BIBforeignlanguage{en}{Data clustering: 50 years beyond
  k-means},'' \emph{\BIBforeignlanguage{en}{Pattern Recognition Letters}},
  vol.~31, no.~8, pp. 651--666, Jun. 2010. [Online]. Available:
  \url{http://linkinghub.elsevier.com/retrieve/pii/S0167865509002323}
\BIBentrySTDinterwordspacing

\bibitem{kriegel_clustering_2009}
\BIBentryALTinterwordspacing
H.-P. Kriegel, P.~Kröger, and A.~Zimek, ``Clustering high-dimensional data: A
  survey on subspace clustering, pattern-based clustering, and correlation
  clustering,'' \emph{{ACM} Trans. Knowl. Discov. Data}, vol.~3, no.~1, pp.
  1:1--1:58, Mar. 2009. [Online]. Available:
  \url{http://doi.acm.org/10.1145/1497577.1497578}
\BIBentrySTDinterwordspacing

\bibitem{macqueen_methods_1967}
\BIBentryALTinterwordspacing
J.~MacQueen, ``\BIBforeignlanguage{{EN}}{Some methods for classification and
  analysis of multivariate observations}.''\hskip 1em plus 0.5em minus
  0.4em\relax The Regents of the University of California, 1967. [Online].
  Available: \url{http://projecteuclid.org/euclid.bsmsp/1200512992}
\BIBentrySTDinterwordspacing

\bibitem{aggarwal_framework_2003}
\BIBentryALTinterwordspacing
C.~C. Aggarwal, J.~Han, J.~Wang, and P.~S. Yu, ``A framework for clustering
  evolving data streams,'' in \emph{Proceedings of the 29th international
  conference on Very large data bases-Volume 29}.\hskip 1em plus 0.5em minus
  0.4em\relax {VLDB} Endowment, 2003, pp. 81--92. [Online]. Available:
  \url{http://dl.acm.org/citation.cfm?id=1315460}
\BIBentrySTDinterwordspacing

\bibitem{aggarwal_framework_2004}
\BIBentryALTinterwordspacing
------, ``A framework for projected clustering of high dimensional data
  streams,'' in \emph{Proceedings of the Thirtieth international conference on
  Very large data bases-Volume 30}.\hskip 1em plus 0.5em minus 0.4em\relax
  {VLDB} Endowment, 2004, pp. 852--863. [Online]. Available:
  \url{http://dl.acm.org/citation.cfm?id=1316763}
\BIBentrySTDinterwordspacing

\bibitem{cao_density-based_2006}
\BIBentryALTinterwordspacing
F.~Cao, M.~Ester, W.~Qian, and A.~Zhou, ``Density-based clustering over an
  evolving data stream with noise.'' in \emph{{SDM}}, vol.~6.\hskip 1em plus
  0.5em minus 0.4em\relax {SIAM}, 2006, pp. 326--337. [Online]. Available:
  \url{http://epubs.siam.org/doi/abs/10.1137/1.9781611972764.29}
\BIBentrySTDinterwordspacing

\bibitem{ntoutsi_density-based_2012}
\BIBentryALTinterwordspacing
I.~Ntoutsi, A.~Zimek, T.~Palpanas, P.~Kröger, and H.-P. Kriegel,
  ``Density-based projected clustering over high dimensional data streams.'' in
  \emph{{SDM}}.\hskip 1em plus 0.5em minus 0.4em\relax {SIAM}, 2012, pp.
  987--998. [Online]. Available:
  \url{http://epubs.siam.org/doi/abs/10.1137/1.9781611972825.85}
\BIBentrySTDinterwordspacing

\bibitem{hassani_subspace_2014}
\BIBentryALTinterwordspacing
M.~Hassani, Y.~Kim, S.~Choi, and T.~Seidl, ``\BIBforeignlanguage{en}{Subspace
  clustering of data streams: new algorithms and effective evaluation
  measures},'' \emph{\BIBforeignlanguage{en}{Journal of Intelligent Information
  Systems}}, Jun. 2014. [Online]. Available:
  \url{http://link.springer.com/10.1007/s10844-014-0319-2}
\BIBentrySTDinterwordspacing

\bibitem{nguyen_survey_2014}
\BIBentryALTinterwordspacing
H.-L. Nguyen, Y.-K. Woon, and W.-K. Ng, ``\BIBforeignlanguage{en}{A survey on
  data stream clustering and classification},''
  \emph{\BIBforeignlanguage{en}{Knowledge and Information Systems}}, Dec. 2014.
  [Online]. Available: \url{http://link.springer.com/10.1007/s10115-014-0808-1}
\BIBentrySTDinterwordspacing

\bibitem{mena-torres_similarity-based_2014}
\BIBentryALTinterwordspacing
D.~Mena-Torres and J.~S. Aguilar-Ruiz, ``\BIBforeignlanguage{en}{A
  similarity-based approach for data stream classification},''
  \emph{\BIBforeignlanguage{en}{Expert Systems with Applications}}, vol.~41,
  no.~9, pp. 4224--4234, Jul. 2014. [Online]. Available:
  \url{http://linkinghub.elsevier.com/retrieve/pii/S0957417413010300}
\BIBentrySTDinterwordspacing

\bibitem{jin_efficient_2014}
\BIBentryALTinterwordspacing
C.~Jin, J.~X. Yu, A.~Zhou, and F.~Cao, ``\BIBforeignlanguage{en}{Efficient
  clustering of uncertain data streams},''
  \emph{\BIBforeignlanguage{en}{Knowledge and Information Systems}}, vol.~40,
  no.~3, pp. 509--539, Sep. 2014. [Online]. Available:
  \url{http://link.springer.com/10.1007/s10115-013-0657-3}
\BIBentrySTDinterwordspacing

\bibitem{liu_clustering_2011}
\BIBentryALTinterwordspacing
W.~Liu and J.~OuYang, ``Clustering algorithm for high dimensional data stream
  over sliding windows.''\hskip 1em plus 0.5em minus 0.4em\relax {IEEE}, Nov.
  2011, pp. 1537--1542. [Online]. Available:
  \url{http://ieeexplore.ieee.org/lpdocs/epic03/wrapper.htm?arnumber=6121009}
\BIBentrySTDinterwordspacing

\bibitem{wan_density-based_2009}
\BIBentryALTinterwordspacing
L.~Wan, W.~K. Ng, X.~H. Dang, P.~S. Yu, and K.~Zhang, ``Density-based
  clustering of data streams at multiple resolutions,'' \emph{{ACM} Trans.
  Knowl. Discov. Data}, vol.~3, no.~3, pp. 14:1--14:28, Jul. 2009. [Online].
  Available: \url{http://doi.acm.org/10.1145/1552303.1552307}
\BIBentrySTDinterwordspacing

\bibitem{lee_efficiently_2009}
\BIBentryALTinterwordspacing
J.~W. Lee, N.~H. Park, and W.~S. Lee, ``\BIBforeignlanguage{en}{Efficiently
  tracing clusters over high-dimensional on-line data streams},''
  \emph{\BIBforeignlanguage{en}{Data \& Knowledge Engineering}}, vol.~68,
  no.~3, pp. 362--379, Mar. 2009. [Online]. Available:
  \url{http://linkinghub.elsevier.com/retrieve/pii/S0169023X0800164X}
\BIBentrySTDinterwordspacing

\bibitem{bellas_model-based_2013}
\BIBentryALTinterwordspacing
A.~Bellas, C.~Bouveyron, M.~Cottrell, and J.~Lacaille,
  ``\BIBforeignlanguage{en}{Model-based clustering of high-dimensional data
  streams with online mixture of probabilistic {PCA}},''
  \emph{\BIBforeignlanguage{en}{Advances in Data Analysis and Classification}},
  vol.~7, no.~3, pp. 281--300, May 2013. [Online]. Available:
  \url{http://link.springer.com/article/10.1007/s11634-013-0133-7}
\BIBentrySTDinterwordspacing

\bibitem{amini_mudi-stream:_2014}
\BIBentryALTinterwordspacing
A.~Amini, H.~Saboohi, T.~Herawan, and T.~Y. Wah,
  ``\BIBforeignlanguage{en}{{MuDi}-stream: A multi density clustering algorithm
  for evolving data stream},'' \emph{\BIBforeignlanguage{en}{Journal of Network
  and Computer Applications}}, Dec. 2014. [Online]. Available:
  \url{http://linkinghub.elsevier.com/retrieve/pii/S1084804514002665}
\BIBentrySTDinterwordspacing

\bibitem{huynh_se-stream:_2014}
\BIBentryALTinterwordspacing
R.~Chairukwattana, T.~Kangkachit, T.~Rakthanmanon, and K.~Waiyamai,
  ``{SE}-stream: Dimension projection for evolution-based clustering of high
  dimensional data streams,'' in \emph{Knowledge and Systems Engineering},
  V.~N. Huynh, T.~Denoeux, D.~H. Tran, A.~C. Le, and S.~B. Pham, Eds.\hskip 1em
  plus 0.5em minus 0.4em\relax Cham: Springer International Publishing, 2014,
  vol. 245, pp. 365--376. [Online]. Available:
  \url{http://link.springer.com/10.1007/978-3-319-02821-7_32}
\BIBentrySTDinterwordspacing

\bibitem{waiyamai_sed-stream:_2014}
\BIBentryALTinterwordspacing
K.~Waiyamai, T.~Kangkachit, T.~Rakthanmanon, and R.~Chairukwattana,
  ``{SED}-stream: Discriminative dimension selection for evolution-based
  clustering of high dimensional data streams,'' \emph{Int. J. Intell. Syst.
  Technol. Appl.}, vol.~13, no.~3, pp. 187--201, Oct. 2014. [Online].
  Available: \url{http://dx.doi.org/10.1504/IJISTA.2014.065174}
\BIBentrySTDinterwordspacing

\bibitem{bohm_density_2004}
C.~Bohm, K.~Railing, H.-P. Kriegel, and P.~Kroger, ``Density connected
  clustering with local subspace preferences,'' in \emph{Fourth {IEEE}
  International Conference on Data Mining, 2004. {ICDM} '04}, Nov. 2004, pp.
  27--34.

\bibitem{_kdd-cup-1999-computer-network-intrusion-detection_????}
\BIBentryALTinterwordspacing
``kdd-cup-1999-computer-network-intrusion-detection.'' [Online]. Available:
  \url{http://www.sigkdd.org/kdd-cup-1999-computer-network-intrusion-detection}
\BIBentrySTDinterwordspacing

\end{thebibliography}
\end{document}